\documentclass[12pt]{article}
\usepackage{subfloat}
\pdfoutput = 1
\usepackage{graphics}
\usepackage{graphicx} 
\usepackage{color}
\usepackage{multirow}
\usepackage{hhline}
\begin{document}
\date{Today}
\title{{\bf{\Large Non-linear effects on the holographic free energy and thermodynamic geometry }}}

\author{
{\bf {\normalsize Debabrata Ghorai}$^{a}$
\thanks{debanuphy123@gmail.com, debabrataghorai@bose.res.in}},\,
{\bf {\normalsize Sunandan Gangopadhyay}$^{b,c}
$\thanks{sunandan.gangopadhyay@gmail.com, sunandan@iiserkol.ac.in,  sunandan@associates.iucaa.in}}\\
$^{a}$ {\normalsize  S.N. Bose National Centre for Basic Sciences,}\\{\normalsize JD Block, 
Sector III, Salt Lake, Kolkata 700098, India}\\[0.2cm]
$^{b}$ {\normalsize Indian Institute of Science Education and Research, Kolkata}\\{\normalsize Mohanpur, Nadia 741246, India}\\ 
$^{c}${\normalsize Visiting Associate in Inter University Centre for Astronomy \& Astrophysics,}\\
{\normalsize Pune 411007, India}\\[0.1cm]
}
\date{}

\maketitle

\begin{abstract}
{\noindent We have analytically investigated the effects of non-linearity on the free energy and thermodynamic geometry of holographic superconductors in $2+1 -$dimensions. The non-linear effect is introduced by considering the coupling of the massive charged scalar field with Born-Infeld electrodynamics. We then calculate the relation between critical temperature and charge density from two different methods, namely, the matching method and the divergence of the scalar curvature which is obtained by investigating the thermodynamic geometry of the model. Both the results indicate a decrease in the critical temperature with increase in the Born-Infeld parameter $b$.  }
\end{abstract}
\vskip 1cm

\section{Introduction}
It is very difficult to study strongly coupled systems using conventional quantum field theory because perturbation technique does not work here. The claim of the AdS/CFT correspondence \cite{adscft1}-\cite{adscft4} is that 4-dimensional strongly coupled gauge theory is related with 5-dimension weakly coupled gravity theory. The theory of strongly coupled systems has been developing recently in modern theoretical physics using this correspondence. The reason for this fascinating development is as follows. This correspondence allows us to map the strongly coupled system to a weakly coupled system. Weakly coupled system can be tackled by perturbation theory and then using this correspondence, one can get a picture of some of the properties of the strongly coupled system which is the topic of interest in our hand.

There has been an upsurge to exploit this duality to explain some of the properties of high $T_c$ superconductors which are wonderful examples of strongly coupled systems. The breakthrough came when it was shown in \cite{hs2} that a scalar hair can form below a certain critical temperature for a charged $AdS$ black hole coupled minimally to a complex scalar field.
The holographic superconductor phase transition \cite{hs3}-\cite{hs6} then involves the scalar operator dual to the scalar field in the bulk acquiring a non-zero vacuum expectation value in the boundary field theory.
A large body of research has also been carried out in the field of holographic fermi liquids \cite{nw1},\cite{nw2}, holographic insulator/superconductor phase transition \cite{nw3}, transport properties of holographic superconductors \cite{hs16}-\cite{dg1}. The machinery which plays a crucial role in achieving this is the AdS/CFT correspondence.

Recently a new direction of research has emerged. It has been realized that the properties of holographic superconductors can be derived from the holographic free energy of such a system \cite{fth2}. In this development a geometrical structure \cite{gs1}-\cite{gs3} is associated with a thermodynamic system in equilibrium. This geometrical structure is obtained from the holographic free energy of the holographic superconductors and can be exploited to obtain the properties of such systems \cite{fth3}. 

In this paper, we exploit the formalism of thermodynamic geometry once again to investigate the properties of $2+1$-dimensional holographic superconductors. However, this time our intention is to study the effects of non-linearity in such systems using this formalism. The non-linearity is introduced by coupling the charged scalar field to Born-Infeld (BI) electrodynamics. There has been a lot of work in which holographic superconductors has been investigated in the presence of BI electrodynamics \cite{jin}-\cite{rb}, \cite{dg2}, \cite{dg1}. It would therefore be interesting to assess these results with the results obtained from the analysis in this work.

To proceed we first obtain the behaviour of the matter fields near the horizon of the black hole using the matching method \cite{siop},\cite{hs8},\cite{fth1},\cite{sgm},\cite{dg3}. The critical temperature and the condensation operator are then computed from the above result.  The study is carried out for the boundary condition $\psi_{+}\neq 0$, $\psi_{-}=0$ for the condensation operator. We then calculate the free energy of this $2+1$-dimensional holographic superconductor in the presence of BI electrodynamics. 
Therefore, the thermodynamic metric computed from this free energy also capture the effects of non-linearity. This is then exploited to calculated the critical temperature from the divergence of the scalar curvature. We finally compare our findings with those obtained from the matching method.

This paper is organized as follows. In section 2, the basic formalism for the holographic superconductors coupled to BI electrodynamics is presented. Section 3 involves the calculation of critical temperature and the condensation operator using the matching method approach. The free energy in terms of the charge density and Born-Infeld parameter is calculated in section 4. In section 5, we calculate the thermodynamic metric and the scalar curvature upto first order in $b$. Section 6 contains the concluding remarks.



\section{Basic formalism }
The plane-symmetric black hole metric can be assumed to take the form
\begin{eqnarray}
ds^2=-f(r)dt^2+\frac{1}{f(r)}dr^2+ r^2 (dx^2 + dy^2)
\label{tg2}
\end{eqnarray}
where $f(r) = r^2 \left( 1 - \frac{r^{3}_{+}}{r^3} \right) $ and $r_{+}$ is the horizon radius. \\
The Hawking temperature of this black hole reads
\begin{eqnarray}
T_{h} = \frac{f^{\prime}(r_{+})}{4\pi} =\frac{3}{4\pi} r_{+} ~.
\label{tg4}
\end{eqnarray} 
This temperature is interpreted as the temperature of the conformal field theory on the boundary. \\
\noindent In $3+1$-dimensions, the action for the model of a holographic superconductor consists a complex scalar field coupled to a $U(1)$ gauge field in anti-de Sitter spacetime 
\begin{eqnarray}
S=\int d^{4}x \sqrt{-g} \left[ \frac{1}{2 \kappa^2} \left( R -2\Lambda \right) -\frac{1}{4} F^{\mu \nu} F_{\mu \nu} -(D_{\mu}\psi)^{*} D^{\mu}\psi-m^2 \psi^{*}\psi \right]
\label{tg0}
\end{eqnarray}
where $\Lambda=-\frac{3}{L^2}$ is the cosmological constant, $\kappa^2 = 8\pi G $, $G$ being the Newton's universal gravitational constant, $A_{\mu}$ and $ \psi $ represent the gauge field and scalar fields. $F_{\mu \nu}=\partial_{\mu}A_{\nu}-\partial_{\nu}A_{\mu}$; ($\mu,\nu=t,r,x,y$) is the field strength tensor, $D_{\mu}\psi=\partial_{\mu}\psi-iqA_{\mu}\psi$ is the covariant derivative.\\
\noindent In this paper, we shall consider Born-Infeld electrodynamics instead of Maxwell electrodynamics for which the actions reads 
\begin{eqnarray}
S=\int d^{4}x \sqrt{-g} \left[ \frac{1}{2 \kappa^2} \left( R -2\Lambda \right) +\frac{1}{b}\left(1-\sqrt{1+\frac{b}{2} F^{\mu \nu} F_{\mu \nu}}\right) -(D_{\mu}\psi)^{*} D^{\mu}\psi-m^2 \psi^{*}\psi \right]
\label{tg1}
\end{eqnarray}
where $b$ is Born-Infeld parameter.\\
\noindent Making the ansatz for the gauge field and the scalar field as \cite{hs3}
\begin{eqnarray}
A_{\mu} = (\phi(r),0,0,0)~,~\psi=\psi(r)
\label{tg3}
\end{eqnarray}
leads to the following equations of motion for the matter fields \cite{sgdc1}, \cite{dg2}
\begin{eqnarray}
\phi^{\prime \prime}(r) + \frac{2}{r} \phi^{\prime}(r) - \frac{2}{r} b \phi^{\prime 3}(r)- \frac{2 q^2 \psi^{2}(r) \phi(r)}{f(r)}\left\{ 1-b\phi^{\prime 2}(r)\right\}^{3/2} = 0
\label{e1}
\end{eqnarray}
\begin{eqnarray}
\psi^{\prime \prime}(r) + \left(\frac{2}{r} + \frac{f^{\prime}(r)}{f(r)}\right)\psi^{\prime}(r) + \left(\frac{q^2 \phi^{2}(r)}{f(r)^2}- \frac{m^{2}}{f(r)}\right)\psi(r) = 0
\label{e01}
\end{eqnarray}
where prime denotes derivative with respect to $r$. We shall also set 
$q=1$ since we shall carry out our analysis in the probe limit \cite{betti}. 
Note that the 
rescaling of the bulk fields $\phi$, $\psi$ and the Born-Infeld
coupling parameter $b$  as $~~\phi \rightarrow \frac{\phi}{q}, ~~\psi \rightarrow \frac{\psi}{q}, ~~ b \rightarrow q^2 b $ puts a factor of $\frac{1}{q^2}$ in front of the matter part of the action (\ref{tg1}). The probe limit corresponds to $\frac{\kappa^2}{q^2}\rightarrow0$ \cite{betti}, \cite{sheykhi}.
For the matter fields to be regular, one requires $\phi(r_+)=0$ and $\psi(r_{+})$ to be finite at the horizon. 

\noindent The matter fields near the boundary of the bulk obey \cite{hs8}
\begin{eqnarray}
\label{bound1}
\phi_{b}(r)&=&\mu-\frac{\rho}{r}\\
\psi_{b}(r)&=&\frac{\psi_{-}}{r^{\Delta_{-}}}+\frac{\psi_{+}}{r^{\Delta_{+}}}
\label{bound2}
\end{eqnarray}
where $ \Delta_{\pm} = \frac{3\pm\sqrt{9+4m^2 L^2}}{2}~.$
The interpretation of the parameters $\mu$ and $\rho$ comes from the gauge/gravity dictionary and are interpreted as the chemical potential and charge density of the conformal field theory on the boundary.

\noindent At this point we make a change of coordinates $z=\frac{1}{r}$. Under this transformation, the metric and the Hawking temperature take the form
\begin{eqnarray}
f(z) &=& \frac{1}{z^2}\left( 1 - \frac{z^3}{z^{3}_{h}} \right) = \frac{1}{z^2} F(z)\\
T_{h} &=& \frac{3}{4\pi z_{h}}
\label{tg5}
\end{eqnarray} 
where $F(z) = (1 - \frac{z^3}{z^{3}_{h}})$ and $z_{h}=\frac{1}{r_+}$~. \\
The field eq.(s) (\ref{e1}),(\ref{e01}) now read
\begin{eqnarray}
\label{tg6}
\phi^{\prime \prime}(z) +2b z^{3}\phi^{\prime 3}(z) - \frac{2 \psi^{2}(z) \phi(z)}{z^2 F(z)}\left\{ 1-b z^{4}\phi^{\prime 2} (z)\right\}^{3/2}  &=& 0 \\
\psi^{\prime \prime}(z) + \left(\frac{F^{\prime}(z)}{F(z)} - \frac{2}{z}\right)\psi^{\prime}(z) + \left(\frac{\phi^{2}(z)}{F^{2}(z)}- \frac{m^{2}}{z^2 F(z)}\right)\psi(z) &=& 0 ~.
\label{tg7}
\end{eqnarray}
In the next section we shall obtain the critical temperature below which the scalar field condensation takes place employing the matching method in the interval $(0, z_h)$. The boundary condition $\phi(r_+)=0$ in $z$-coordinate translates to $\phi(z=z_h)=0$. The asymptotic behaviour of the fields read
\begin{eqnarray}
\phi_{b} (z) &=& \mu - \rho z \\
\psi_{b} (z) &=& \psi_{-} z^{\Delta_{-}}~ +~ \psi_{+} z^{\Delta_{+}}~~.
\label{tg8}
\end{eqnarray}



\section{Critical temperature from the matching method}
\noindent The Taylor series expansions of the fields near the horizon read
\begin{eqnarray}
\label{tg9}
\phi_{h}(z) &=& \phi(z_h ) + \phi^{\prime}(z_h)(z-z_h) + \frac{\phi^{\prime \prime}(z_h)}{2}(z-z_h)^2 + ..........\\
\psi_{h}(z) &=& \psi(z_h ) + \psi^{\prime}(z_h)(z-z_h) + \frac{\psi^{\prime \prime}(z_h)}{2}(z-z_h)^2 + .........
\label{tg10}
\end{eqnarray} 
We shall now determine the undetermined coefficients using the boundary condition $\phi(z_h) = 0$ along with $f(z_h) = 0$ and eq.(s)(\ref{tg6}),(\ref{tg7}). This yields upto first order in the Born-Infeld parameter
\begin{eqnarray}
\label{tg11}
\phi^{\prime\prime}(z_h) &=& -\left[2b z^{3}_{h}\phi^{\prime 3}(z_{h}) +\frac{2 \phi^{\prime}(z_h) \psi^{2}(z_h)}{3 z_h}\left(1-\frac{3}{2}b z^{4}_{h}\phi^{\prime 2}(z_{h} )\right)\right] \\
\psi^{\prime}(z_h) &=& -\frac{m^2}{3 z_h} \psi(z_h)~~~ ; ~~~\psi^{\prime\prime}(z_h) = \frac{\psi(z_h)}{18 z^{2}_{h}}\left[ m^4 + 6 m^2 - z^4 \phi^{\prime 2}(z_h) \right].
\label{tg12}
\end{eqnarray}
\noindent  This is consistent with the Breitenlohner-Freedman bound \cite{fth4},\cite{bf2}. Hence the near horizon expansions of these fields upto $\mathcal{O}((z-z_{h})^2)$ read 
\begin{eqnarray}
\label{tg13}
\phi_{h}(z) &=& \phi^{\prime}(z_h) \left[(z-z_h) - \frac{\psi^{2}(z_h)}{3 z_h}(z-z_h)^{2} \right] +b\phi^{\prime 3}(z_h) z^{3}_{h}(z-z_{h})^{2}\left[\frac{\psi^{2}(z_{h})}{2} -1\right] \\
\psi_{h}(z) &=& \psi(z_h) \left[1 - \frac{m^2}{3 z_h}(z - z_h) + \frac{m^4 +6m^2 - z^{4}_{h} \phi^{\prime 2}(z_h) }{36 z^{2}_{h}}(z-z_h)^{2} \right]~.
\label{tg14}
\end{eqnarray}
We now proceed to match the near horizon expression of the fields with the asymptotic solution of these field at any arbitrary point between the horizon and the boundary, say $z=\frac{z_{h}}{2}$. We would like to make a comment at this point. In the investigation carried out in \cite{dg3}, we matched the  near horizon and the asymptotic solutions at any arbitrary point between the horizon and the boundary, that is at $z=\frac{z_h}{\lambda}$ with $\lambda$ lying between $[1, \infty]$. We observed that the agreement between the matching method values and the numerical values was very good for $\lambda =3$. However, there is no way to determine a priori a proper matching point. One can in principle choose other matching points also. Thus, the matching method is technically ambiguous compared to the Sturm-Liouville eigenvalue method. In this paper, we choose $\lambda=2 $ for simplicity.

\noindent The matching conditions are 
\begin{eqnarray}
\label{tg15}
\phi_{h}\left(\frac{z_h}{2}\right) = \phi_{b}\left(\frac{z_h}{2}\right)  ~~~;~~~  \phi^{\prime}_{h}\left(\frac{z_h}{2}\right) &=& \phi^{\prime}_{b}\left(\frac{z_h}{2}\right) \\
\psi_{h}\left(\frac{z_h}{2}\right) = \psi_{b}\left(\frac{z_h}{2}\right)  ~~~;~~~  \psi^{\prime}_{h}\left(\frac{z_h}{2}\right) &=& \psi^{\prime}_{b}\left(\frac{z_h}{2}\right).
\label{tg16}
\end{eqnarray} 
From eq.(\ref{tg15}), we obtain the following relations
\begin{eqnarray}
\label{tg17a}
\psi^{2}(z_{h}) &=& -\frac{4}{[1 -\frac{3b}{2}z^{4}_{h}\phi^{\prime 2}(z_{h})]} \left( \frac{\mu}{z_{h} \phi^{\prime}(z_h)} + 1 + \frac{3b}{4}z^{4}_{h}\phi^{\prime 2}(z_{h}) \right)  \\
\rho &=& -\phi^{\prime}(z_{h})\left[1+ \frac{\psi^{2}(z_{h})}{3}\right] +b\phi^{\prime 3}(z_{h}) z^{4}_{h}\left[\frac{\psi^{2}(z_{h})}{2}- 1\right]~.
\label{tg17}
\end{eqnarray}
From eq.(\ref{tg16}), we get 
\begin{eqnarray}
\label{tg18}
\psi_{-/+} &=& \frac{m^2 +12}{6+3\Delta}.\frac{2^{\Delta- 1}}{z^{\Delta}_{h}}.\psi(z_{h}) \\
\phi^{\prime 2}(z_h) &=& \frac{1}{z^{4}_{h}}\left[ \frac{144\Delta +6m^{2}(6+5\Delta)+m^{4}(2+\Delta)}{2+\Delta} \right] ~~\Rightarrow ~ \phi^{\prime}(z_{h}) = -\frac{\chi(m,\Delta)}{z^{2}_{h}}
\label{tg19}
\end{eqnarray} 
where 
\begin{eqnarray}
\chi(m,\Delta) = \sqrt{\frac{144\Delta +6m^{2}(6+5\Delta)+m^{4}(2+\Delta)}{2+\Delta} }~.
\end{eqnarray}
\noindent We shall set $m^2 = -2 ~$in the rest of our analysis.
In eq.(\ref{tg18}), $\psi_{-}$ is for $\Delta= \Delta_{-} = 1 $ and $\psi_{+}$ is for $\Delta =\Delta_{+} = 2$.\\
\noindent We now substitute $\phi^{\prime}(z_h)$ from eq.(\ref{tg19}) in eq.(s)(\ref{tg17a}),(\ref{tg17}) to obtain upto first order in $b$
\begin{eqnarray}
\psi(z_h) &=& 2\sqrt{\left(1+\frac{3b}{2}\chi^{2}\right)\frac{\mu z_{h}}{\chi} -\left(1+\frac{9b}{4}\chi^{2}\right)} \\
 &=& \sqrt{3}\sqrt{\left(1+\frac{3b}{2}\chi^{2}\right)\frac{\rho z^{2}_{h}}{\chi} -\left(1+\frac{5b}{2}\chi^{2}\right)}\\
\mu &=& \frac{\chi}{z_h} \left[ \frac{3\rho {z_{h}^{2}}}{4\chi} + \frac{1}{4}\right].
\label{rho1}
\end{eqnarray} 
The condensation operator and the critical temperature in terms of the chemical potential and the charge density can now be obtained using eq.(s)(\ref{tg17a})-(\ref{tg19}) and $T = \frac{3}{4\pi z_h}$ : 
\begin{eqnarray}
\langle \mathcal{O}\rangle = \gamma_{(\mu)} T^{\Delta}_{c} \left( 1 - \frac{T}{T_{c}}\right)^{1/2} ~~~ ; ~~~ T_{c} &=& \xi_{(\mu)} \mu \\
\langle \mathcal{O}\rangle = \gamma_{(\rho)} T^{\Delta}_{c} \left( 1 - \frac{T}{T_{c}}\right)^{1/2} ~~~ ; ~~~ T_{c} &=& \xi_{(\rho)} \sqrt{\rho}
\end{eqnarray}
where
\begin{eqnarray}
\gamma_{(\mu)} = \frac{(m^2 + 12) 2^{\Delta}\sqrt{1+\frac{9b}{4}\chi^2} }{(6+3\Delta)}\left(\frac{4\pi}{3}\right)^{\Delta} ~~;~~ \xi_{(\rho)} &=& \frac{1+\frac{3b}{2}\chi^{2}}{1+\frac{9b}{4}\chi^{2}} \frac{3}{4\pi\chi}~~\\
\gamma_{(\rho)} = \frac{(m^2 + 12) 2^{\Delta -1}\sqrt{6(1+\frac{5b}{2}\chi^2)} }{(6+3\Delta)}\left(\frac{4\pi}{3}\right)^{\Delta} ~~;~~ \xi_{(\rho)} &=& \sqrt{\frac{1+\frac{3b}{2}\chi^{2}}{1+\frac{5b}{2}\chi^{2}}} \frac{3}{4\pi\sqrt{\chi}}~~.
\end{eqnarray}

\noindent We now consider $\psi_{+}= \langle \mathcal{O}\rangle, ~ \psi_{-}= 0$ which implies that the condensate $\psi_{+}$ forms in the absence of the source term $\psi_{-}$. Hence the condensation operator and the critical temperature in terms of the chemical potential and the charge density read
\begin{eqnarray}
\langle \mathcal{O}_{+}\rangle_{\mu} = \frac{160\pi^{2}\sqrt{1+ 63 b} }{27 }T^{2}_{c} \left( 1 - \frac{T}{T_{c}}\right)^{1/2}~~;~~ T_{c} &=& 0.0451 \left(\frac{1+42 b}{1+63 b}\right) \mu 
\end{eqnarray}
\begin{eqnarray}
\langle \mathcal{O}_{+}\rangle_{\rho} = \frac{80\pi^{2}\sqrt{6(1+ 70b)} }{27 }T^{2}_{c} \left( 1 - \frac{T}{T_{c}}\right)^{1/2}~~;~~ T_{c} &=& 0.104  \sqrt{\frac{1+42 b}{1+70 b}} \sqrt{\rho}~.
\end{eqnarray}\\
In Table \ref{E4}, we have presented the analytical values of the coefficients of the $\mu$ (in the $T_{c}-\mu$ relation) for different values of $b$ obtained from the matching method.


\section{Free energy of the non-linear holographic superconductor}  
In this section we shall calculate the free energy at a finite temperature of the field theory living on the boundary of the $3+1$- bulk theory. Here we essentially follow the technique in \cite{dg3} .\\
The first step is to consider the action for the action for the Abelian-Higgs sector
\begin{eqnarray}
S_{M} = \int d^{4}x \sqrt{-g} \left[\frac{1}{b}\left(1-\sqrt{1+\frac{b}{2} F^{\mu \nu} F_{\mu \nu}}\right) -(D_{\mu}\psi)^{*} D^{\mu}\psi-m^2 \psi^{*}\psi \right]~.
\end{eqnarray}
Setting $m^{2}=-2$ and expanding the above action keeping terms upto linear order in $b$ yields
\begin{eqnarray}
S_{M}= \int d^{4}x \left[\frac{\phi^{\prime 2}(z)}{2} - \frac{F(z)\psi^{\prime 2}(z)}{z^{2}} + \frac{\phi^{2}(z)\psi^{2}(z)}{z^{2}F(z)} +\frac{2\psi^{2}(z)}{z^4} + b \frac{z^4 \phi^{\prime 4}(z)}{8}+ \mathcal{O} (b^{2}) \right]~.
\end{eqnarray}
Applying the boundary condition ($\phi(z_{h}) = 0$) and the equations of motion (\ref{tg6}),(\ref{tg7}) the on-shell value of the action $S_{E}$ reads
\begin{eqnarray}
S_{o} &=& \int d^{3}x \left[-\frac{1}{2}\phi (z)\phi^{\prime}(z) \mid_{z=0}~ + \frac{F(z)\psi(z)\psi^{\prime}(z)}{z^{2}}\mid_{z=0} \right. \nonumber \\
&& \left. - \int^{z_{h}}_{0} dz \left( \frac{\phi^{2}(z)\psi^{2}(z)}{z^{2}F(z)} -\frac{3b}{4}\frac{z^2 \psi^{2}(z)\phi^{2}(z)\phi^{\prime 2}(z)}{F(z)} -\frac{b}{2} z^3 \phi(z)\phi^{\prime 3}(z) \right)\right] ~.
\label{alexander}
\end{eqnarray}
The asymptotic behaviour of $\phi(z)$ and $\psi(z)$ is now substituted in the above expression. This gives
\begin{eqnarray}
S_{o} &=& \int d^{3}x \left[\frac{\mu\rho}{2} + 3\psi_{+}\psi_{-} +\left(\frac{\psi^{2}_{-}}{z}\right)\mid_{z=0} \right. \nonumber \\
&& \left. ~-~ \int^{z_{h}}_{0} dz \left( \frac{\phi^{2}(z)\psi^{2}(z)}{z^{2}(1-z^{3}/z^{3}_{h})} -\frac{3b}{4}\frac{z^2 \psi^{2}(z)\phi^{2}(z)\phi^{\prime 2}(z)}{1-z^{3}/z^{3}_{h}} -\frac{b}{2} z^3 \phi(z)\phi^{\prime 3}(z) \right)\right]~.
\end{eqnarray}
From the above form of the on-shell action, we observe that the term $\left(\frac{\psi^{2}_{-}}{z}\right)\mid_{z=0}$ diverges. We study the holographic free energy for the boundary condition $\psi_{+}= \langle \mathcal{O}\rangle,~\psi_{-}=0$. To cancel this divergence we add a counter term at the boundary. The counter action which gives the counter term required to cancel this divergence reads
\begin{eqnarray}
S_{c} = -\int d^{3}x \left(\sqrt{-h} \psi^{2}(z) \right)\mid_{z=0}
\end{eqnarray} 
where $h$ is the determinant of the induced metric on the $AdS$ boundary. We now evaluate this using the asymptotic behaviour of $\psi(z)$ to obtain
\begin{eqnarray}
S_{c} = -\int d^{3}x \left[2\psi_{+}\psi_{-} + \left(\frac{\psi^{2}_{-}}{z}\right)\mid_{z=0} \right]~.
\end{eqnarray}\\
Adding $S_o$ and $S_c$ leads to The free energy of the $2+1$-boundary field theory. 
\begin{eqnarray}
\Omega = -T (S_{o} + S_{c}) = \beta T V_{2} \left[ -\frac{\mu\rho}{2} -\psi_{+}\psi_{-} + I~\right] 
\end{eqnarray}
where $\int d^{3}x = \beta V_{2}$, $V_{2}$ being the volume of the $2-$dimensional space of the boundary and the integral $I$ reads upto first order in the BI parameter $b$
\begin{eqnarray}
I &=& \int^{z_h}_{0} dz \frac{\phi^{2}(z)\psi^{2}(z)}{z^{2}(1- z^{3}/z^{3}_{h})} -\frac{3b}{4} \int^{z_h}_{0} dz \frac{z^2 \psi^{2}(z)\phi^{2}(z)\phi^{\prime 2}(z)}{1-z^{3}/z^{3}_{h}} 
-b \int^{z_h}_{0} dz z^{3}\phi^{\prime 3}(z)\left[ \phi (z) +\frac{z \phi^{\prime}(z)}{8}\right] \nonumber \\
&=& \int^{z_{h}/2}_{0} dz \frac{\phi^{2}_{b}(z)\psi^{2}_{b}(z)}{z^{2}(1- z^{3}/z^{3}_{h})} + \int^{z_h}_{z_{h}/2} dz \frac{\phi^{2}_{h}(z)\psi^{2}_{h}(z)}{z^{2}(1- z^{3}/z^{3}_{h})} -\frac{3b}{4} \int^{z_{h}/2}_{0} dz \frac{z^2 \psi^{2}_{b}(z)\phi^{2}_{b}(z)\phi^{\prime 2}_{b}(z)}{1-z^{3}/z^{3}_{h}} \nonumber \\
&-& \frac{3b}{4} \int^{z_h}_{z_{h}/2} dz \frac{z^2 \psi^{2}_{h}(z)\phi^{2}_{h}(z)\phi^{\prime 2}_{h}(z)}{1-z^{3}/z^{3}_{h}} - \frac{b}{2} \int^{z_h /2}_{0} dz z^{3}\phi_{b} (z)\phi^{\prime 3}_{b} (z)  -\frac{b}{2} \int^{z_h}_{z_{h}/2} dz z^{3}\phi_{h} (z)\phi^{\prime 3}_{h} (z).
\end{eqnarray}

\noindent Following the technique in \cite{dg3} to evaluate the above integral and substituting the value of $\rho$ in terms of $\mu$ from eq.(\ref{rho1}) and using $z_h =\frac{3}{4\pi T}$, we obtain the expression for the free energy in terms of the chemical potential
\begin{eqnarray}
\frac{\Omega}{V_2}  &=& -\frac{\mu\rho}{2} + I \nonumber \\
& = & E_{1} T^3 + E_{2} T^{2}\mu + E_{3} T \mu^{2} +  E_{4}\mu^{3} + b \left[ E_{5} T^{3} + E_{6} T^{2} \mu + E_{7} T \mu^{2} +  E_{8}\mu^{3} + E_{9}\frac{\mu^4}{T} +  E_{10}\frac{\mu^5}{T^2} \right] \nonumber \\
\end{eqnarray}
where $E_{i}~(i=1,2,...,10)$ are constants.\\
Substituting the value of $\mu$ in terms of $\rho$ from eq.(\ref{rho1}), we obtain the expression for the free energy in terms of the charge density 
\begin{eqnarray}
\frac{\Omega}{V_2} = F_{1} T^3 + F_{2}\rho T + F_{3}\frac{\rho^2}{T} +  F_{4}\frac{\rho^3}{T^3} + b \left[ F_{5} T^{3} + F_{6}\rho T + F_{7}\frac{\rho^2}{T} +  F_{8}\frac{\rho^3}{T^3} + F_{9}\frac{\rho^4}{T^5} +  F_{10}\frac{\rho^5}{T^7} \right] \nonumber \\
\end{eqnarray}
where $F_{i}~(i=1,2,...,10)$ are constants.\\
In Table \ref{t2}, we have shown the values of these constants for the boundary condition $\psi_{+}= \langle \mathcal{O}\rangle,~\psi_{-}=0$.
\begin{table}[h!]
\caption{The values of $E_{i},~F_{i},~(i=1,2,...,10)$ for $\Delta=\Delta_{+}=2$.}
\centering
\begin{tabular}{|c| c| c| c| c| c| c| c| c| c|}
\hline
$E_{1}$ & $E_{2}$ & $E_{3}$ & $E_{4}$ & $E_{5}$ & $E_{6}$ & $E_{7}$ & $E_{8}$ & $E_{9}$ & $E_{10}$ \\
\hline
-167.89 & 8.407 & -2.882 & 0.0338 & -10154.2 & -587.24 & -18.989 & 1.434 & -0.0030 & 0.000089 \\
\hline
\hline
$F_{1}$ & $F_{2}$ & $F_{3}$ & $F_{4}$ & $F_{5}$ & $F_{6}$ & $F_{7}$ & $F_{8}$ & $F_{9}$ & $F_{10}$ \\
\hline
-204.04 & -3.655 & -0.0744 & 0.00019 & -13745 & -118.89 & 0.1683 & 0.0085 & $5.4\times10^{-6}$ & -$ 1.6\times 10^{-8}$ \\
\hline
\end{tabular}
\label{t2}
\end{table} 

\section{Thermodynamic geometry}
\noindent In this section we move on to investigate the thermodynamic geometry of the holographic superconductor. The thermodynamic metric is defined as \cite{gs1}\footnote{Note that the on-shell action (\ref{alexander}) in our analysis with proper counter term added to it is identified (upto a volume factor) with the free energy $\omega (T,\mu)$ in the grand canonical ensemble. Hence, $\omega (T,\rho)$ should be understood as $\omega (T,\rho(\mu))$ with $x^i = (T,\mu)$. However, the formula 
$g_{ij} = -\frac{1}{T}\frac{\partial^2 \omega(T,\rho)}{\partial x^{i} \partial x^{j}}$
with $x^i = (T,\rho)$ can also be used to compute the thermodynamic geometry and holds in the canonical ensemble. Both the formulas for the thermodynamic metric are equivalent to each other and are related by Legendre transformation.}
\begin{eqnarray}
g_{ij} = -\frac{1}{T}\frac{\partial^2 \omega(T,\mu)}{\partial x^{i} \partial x^{j}}
\label{tgm1}
\end{eqnarray}
where $\omega = \frac{\Omega}{V_{2}},~ x^1 = T $ and $x^2= \mu~$.
Hence the components of the metric in terms of $\mu$ upto first order in $b$ read
\begin{eqnarray}
g_{TT} &=& -\left[6 E_{1} + 2 E_{2}\frac{\mu}{T} + b\left( 6 E_{5} +2E_{6}\frac{\mu}{T}+ 2E_{9}\frac{\mu^4}{T^4} + 6 E_{10}\frac{\mu^5}{T^5}  \right) \right]  \nonumber \\
g_{T\mu} &=& g_{\mu T} =-\left[2E_{2} + 2 F_{3}\frac{\mu}{T} + b\left( 2E_{6} + 2 E_{7}\frac{\mu}{T} - 4E_{9}\frac{\mu^3}{T^3} - 10 E_{10}\frac{\mu^4}{T^4} \right)\right] \nonumber \\
g_{\mu\mu} &=& -\left[2E_{3} + 6 E_{4}\frac{\mu}{T} +b\left( 2E_{7}  +6E_{8}\frac{\mu}{T} + 12E_{9}\frac{\mu^2}{T^2} +20E_{10}\frac{\mu^3}{T^3}\right)\right]~.
\label{tg59a}
\end{eqnarray}
The scalar curvature of a general metric
\begin{eqnarray}
ds^{2}_{th} = g_{11} (dx^{1})^{2} + 2 g_{12} dx^{1} dx^{2} + g_{22} (dx^{2})^2
\label{tg60}
\end{eqnarray}
is given by \cite{gs2}
\begin{eqnarray}
R =\frac{-1}{\sqrt{g}}\left[ \frac{\partial}{\partial x^{1}}\left(\frac{g_{12}}{g_{11}\sqrt{g}}\frac{\partial g_{11}}{\partial x^{2}} - \frac{1}{\sqrt{g}}\frac{\partial g_{22}}{\partial x^{1}}\right) + \frac{\partial}{\partial x^{2}}\left(\frac{2}{\sqrt{g}}\frac{\partial g_{22}}{\partial x^{2}} - \frac{1}{\sqrt{g}}\frac{\partial g_{11}}{\partial x^{2}} -\frac{g_{12}}{g_{11}\sqrt{g}}\frac{\partial g_{11}}{\partial x^{1}}\right) \right]~.
\label{tg61}
\end{eqnarray} 
A singularity in $R$ can be found by checking whether the denominator of the right hand side of eq.(\ref{tg61}) vanishes. The condition of the divergence of $R$ is $ det g_{ij} = 0$ which reads
\begin{eqnarray}
g_{TT} g_{\mu\mu} - g^{2}_{T\mu} = 0~~.
\end{eqnarray}
The temperature for which the scalar curvature diverges can be obtained by solving this equations keeping terms upto first order in $b$. This temperature is said to the critical temperature in the formalism of thermodynamic geometry. We obtain this critical temperature for the case $\psi_{+}\neq 0~,~\psi_{-}= 0~$ for different values of $b$ and compare them with the results which have been obtained from the matching method. \\
\noindent Table \ref{E4} gives the results for $\Delta=\Delta_{+}=2$ obtained from the matching method and the thermodynamic geometry for a set of values of the Born-Infeld parameter $b$.
\begin{table}[ht]
\caption{ The critical temperature $T_{c}=\xi_{(\mu)}\mu $ 
for $\Delta = \Delta_{+} =2$.}   
\centering                          
\begin{tabular}{|c| c| c|}            
\hline
$b$ & $\xi_{(\mu)}$ from Matching Method & $\xi_{(\mu)}$ from divergence of $R$ \\
\hline
 0.0 & 0.0451 & 0.0838 \\ 
\hline
 0.01  & 0.0393 & 0.0823 \\
\hline
 0.02 & 0.0367 & 0.0815  \\
\hline 
0.03 & 0.0329 & 0.0811 \\
\hline          
\end{tabular}
\label{E4}  
\end{table}



\section{Conclusions}
We now list our results obtained in this study. We obtain the value of the critical temperature and the condensation operator of a holographic superconductor coupled to Born-Infeld electrodynamics living in a $2+1$-dimensions using the formalism of thermodynamic geometry and matching method.  The results from the matching method show that the critical temperature decreases with increase in the Born-Infeld parameter. We observe that the thermodynamic geometry gets affected due to the Born-Infeld parameter which in turn makes the critical temperature to depend on this parameter. The computation of the free energy of the holographic superconductor is carried out in the BI framework and is observed to capture the effects of non-linearity. This then leads to the thermodynamic geometry. Here also we observe that there is a decrease in the critical temperature when the BI parameter $b$ increased. This tells that the condensation gets harder to form with increase in non-linearity.

\section*{Acknowledgments} DG would like to thank DST-INSPIRE, Govt. of India for financial support. DG would also like to thank Ankan Pandey of S.N.Bose Centre for helping in Mathematica. SG acknowledges the support by DST SERB under Start Up Research Grant (Young Scientist), File No.YSS/2014/000180. The authors thank the referee for useful comments.


\end{document}